\documentclass[12pt]{article}
\usepackage{array,amsmath,amssymb,epsfig,verbatim}
\begin{document}


\newcommand{\kIR}{K_{\Lambda\infty}}
\newcommand{\kUV}{K_{0\Lambda}}
\newcommand{\Str}{\mbox{STr}}

\def\slash#1{\ooalign{$\hfil/\hfil$\crcr$#1$}}
\def\KM#1{K_{\Lambda\Lambda_0}(-(#1)^2)} 

\newcommand{\bracket}[2]
{
\left[ #1 \right]_{#2}
}
\newcommand{\sezione}[1]
{
{$\ $ \\ \bf\large #1 \\ } \addcontentsline{toc}{section}{#1}
}


\def\ap#1#2#3{Ann.\ Phys.\ (NY) #1 (19#3) #2}
\def\cmp#1#2#3{Commun.\ Math.\ Phys.\ #1 (19#3) #2}
\def\ib#1#2#3{ibid.\ #1 (19#3) #2}
\def\zp#1#2#3{Z.\ Phys.\ #1 (19#3) #2}
\def\np#1#2#3{Nucl.\ Phys.\ B#1 (19#3) #2}
\def\pl#1#2#3{Phys.\ Lett.\ #1B (19#3) #2}
\def\pr#1#2#3{Phys.\ Rev.\ D #1 (19#3) #2}
\def\prb#1#2#3{Phys.\ Rev.\ B #1 (19#3) #2}
\def\prep#1#2#3{Phys.\ Rep.\ #1 (19#3) #2}
\def\prl#1#2#3{Phys.\ Rev.\ Lett.\ #1 (19#3) #2}
\def\rmp#1#2#3{Rev.\ Mod.\ Phys.\ #1 (19#3) #2}

\newcommand{\figura}[4]
{
\begin{figure}[t]
  \begin{center}
  \mbox{\epsfig{file=#1,height=#2}}
  \end{center}
  \label{#3}
  \caption{{{\small #4}}}
\end{figure}
}
\newcommand{\figuraX}[2]
{
\begin{figure}[h]
  \begin{center}
  \mbox{\epsfig{file=#1,height=#2}}
  \end{center}
\end{figure}
}
\renewcommand{\div}
{
\mbox{div}
}
\newcommand{\binomio}[2]
{
\pmatrix{#1\cr #2}
}
\newcommand{\disegno}[2]
{
\begin{center}                             
  \begin{picture}(#1) #2
  \end{picture}
\end{center}
}
\newcommand{\vettore}[3]
{
\put(#1){\vector(#2){#3}}
}
\newcommand{\entra}[4]
{
\put(#1){\makebox(0,0){$\boldmath\times$}}
\put(#1){\line(#2){#3}}
\put(#1){\makebox(0,-4){$#4$}}
}
\newcommand{\entraX}[4]
{
\put(#1){\makebox(0,0){$\boldmath\times$}}
\put(#1){\line(#2){#3}}
\put(#1){\makebox(-3,-2){$#4$}}
}
\newcommand{\scrivi}[2]
{
\put(#1){\makebox(0,0){$#2$}}
}
\newcommand{\blob}[2]
{
\put(#1){\circle{2}}
\put(#1){\makebox(-.5,-.5){$#2$}} 
}                    
\newcommand{\croce}[2]
{
\put(#1){\makebox(0,0){$\boldmath\times$}}
\put(#1){\makebox(-3,-2){$#2$}} 
}                       
\newcommand{\puntino}[2]
{
\put(#1){\circle*{.5}}
\put(#1){\makebox(0,-4){$#2$}} 
}                 
\newcommand{\puntinoX}[2]
{
\put(#1){\circle*{.5}}
\put(#1){\makebox(-3,-2){$#2$}} 
}                       
\newcommand{\punto}[2]
{
\put(#1){\circle*{.9}}
\put(#1){\makebox(0,-4){$#2$}} 
}
\newcommand{\puntoX}[2]
{
\put(#1){\circle*{.9}}
\put(#1){\makebox(-3,-2){$#2$}} 
}
\newcommand{\puntone}[2]
{
\put(#1){\circle*{1.5}}
\put(#1){\makebox(0,-4){$#2$}} 
}
\newcommand{\puntoneX}[2]
{
\put(#1){\circle*{1.5}}
\put(#1){\makebox(-3,-2){$#2$}} 
}
\newcommand{\acapo}
{
\\ \ \\ 
}
\newcommand{\gr}[1]
{
^{\underline{#1}} 
}
\newcommand{\grr}[2]
{
^{\underline{#1\times #2}} 
}
\newcommand{\des}[1]
{
\overleftarrow{#1}
}
\newcommand{\ddes}[2]
{
{\overleftarrow\delta #1\over\delta #2}
}
\newcommand{\dsin}[2]
{
{\overrightarrow\delta #1\over\delta #2}
}
\newcommand{\valuta}[1]
{
\left.#1\right|
}

\newcommand{\iacc}{\`\i{\ }}
\renewcommand{\a}{{\tilde a}} \renewcommand{\b}{{\tilde b}}
\renewcommand{\c}{{\tilde c}} \renewcommand{\d}{{\tilde d}}
\newcommand{\e}{{\tilde e}}\newcommand{\f}{{\tilde f}}
\newcommand{\g}{{\tilde g}}\newcommand{\h}{{\tilde h}}
\newcommand{\ii}{{\tilde \imath}} \newcommand{\jj}{{\tilde \jmath}}
\renewcommand{\k}{{\tilde k}} \renewcommand{\l}{{\tilde l}}
\newcommand{\m}{{\tilde m}} \newcommand{\n}{{\tilde n}}
\renewcommand{\o}{{\tilde o}} \newcommand{\p}{{\tilde p}}
\newcommand{\q}{{\tilde q}} \renewcommand{\r}{{\tilde r}}
\newcommand{\s}{{\tilde s}} \renewcommand{\t}{{\tilde t}}
\renewcommand{\u}{{\tilde u}} \renewcommand{\v}{{\tilde v}}
\newcommand{\w}{{\tilde w}} \newcommand{\x}{{\tilde x}}
\newcommand{\y}{{\tilde y}} \newcommand{\z}{{\tilde z}}
\newcommand{\ux}{{\underline x}} \newcommand{\uy}{{\underline y}}
\newcommand{\uv}{{\underline v}} \newcommand{\uw}{{\underline w}}
\newcommand{\uz}{{\underline z}}
\newcommand{\bi}{\bar\imath} \newcommand{\bj}{\bar\jmath}


\newcommand{\tadpole}{\mbox{ 
\begin{picture}(12,4)
  \put(0,1){\line(1,0){10}}
  \put(5,2.5){\circle{3}}
\end{picture}}}

\newcommand{\zetadue}{\mbox{
\begin{picture}(6,3)
  \put(0,1){\line(1,0){6}}
  \put(3,1){\circle{2}}
\end{picture}}}

\newcommand{\vertice}{\mbox{ 
\begin{picture}(14,6)
 \put(6,2){\circle{8}}
 \put(2,2){\line(-1,1){4}}
 \put(2,2){\line(-1,-1){4}}
 \put(10,2){\line(1,1){4}}
 \put(10,2){\line(1,-1){4}}
\end{picture}}}

\newcommand{\verticeD}{\mbox{ 
\begin{picture}(8,3)
  \put(1,2){\line(1,0){6}}
  \put(4,2){\circle{2}}
  \put(4,1){\line(-1,-1){1.5}}
  \put(4,1){\line(1,-1){1.5}}
\end{picture}}}

\newcommand{\doppiovert}{\mbox{ 
\begin{picture}(6,3)
  \put(1,1){\line(-1,1){1.5}}
  \put(1,1){\line(-1,-1){1.5}}
  \put(2,1){\circle{2}}
  \put(4,1){\circle{2}}
  \put(5,1){\line(1,1){1.5}}
  \put(5,1){\line(1,-1){1.5}}
\end{picture}}}

\newcommand{\doppioquatt}{\mbox{ 
\begin{picture}(6,3)
  \put(2,1){\line(-1,1){1.5}}
  \put(2,1){\line(-1,-1){1.5}}
  \puntino{5,1}{}
  \put(3.5,1){\circle{3}}
  \put(5,1){\line(1,1){1.5}}
  \put(5,1){\line(1,-1){1.5}}
\end{picture}}}

\newcommand{\verticesei}{\mbox{
\begin{picture}(14,6)
 \put(6,2){\circle{8}}
 \put(6,6){\line(1,1){4}}
 \put(6,6){\line(-1,1){4}}
 \put(2,2){\line(-1,1){4}}
 \put(2,2){\line(-1,-1){4}}
 \put(10,2){\line(1,1){4}}
 \put(10,2){\line(1,-1){4}}
\end{picture}}}

\newcommand{\ctdue}{\mbox{
\begin{picture}(12,2)
  \put(0,1){\line(1,0){10}} \puntone{5,1}{}
\end{picture}}}

\newcommand{\ctquatt}{\mbox{
\begin{picture}(12,2)
  \put(0,1){\line(1,0){10}}
  \puntone{5,1}{}
  \put(5,1){\line(1,-1){4}}
  \put(5,1){\line(-1,-1){4}}
\end{picture}}}
\def\bom#1{\mbox{\boldmath$#1$}}

\def\ldl{\Lambda_0\frac{\partial}{\partial\Lambda_0}}
\def\mdm{\mu\frac\partial{\partial\mu}}
\def\lDl{\Lambda_0\frac{d}{d\Lambda_0}} 

\def\LdL{\Lambda\partial_\Lambda}
\def\LDL{\Lambda \delta_\Lambda}

\def\UV{ {\mbox{\scriptsize UV}} }
\def\limUV{\lim_{\Lambda_0\to\infty}}
\def\MSbar{ \overline{MS} }
\def\arctanh{\mbox{arctanh}}


\newcommand{\hor}
  {\mbox{hor}}
\newcommand{\ver}
  {\mbox{ver}}
\newcommand{\Tr}
  {\mbox{Tr}}
\newcommand{\STr}
  {\mbox{STr}}
\newcommand{\A}
  {{\cal A}}
\newcommand{\B}
  {{\cal B}}
\newcommand{\C}
  {{\cal C}}                                  
\newcommand{\D}
  {{\cal D}}
\newcommand{\De}
  {{\cal D}}
\newcommand{\E}
  {{\cal E}}
\newcommand{\F}
  {{\cal F}}
\newcommand{\G}
  {{\cal G}}
\renewcommand{\H}
  {{\cal H}}
\newcommand{\I}
  {{\cal I}}
\newcommand{\J}
  {{\cal J}}
\newcommand{\K}
  {{\cal K}}
\renewcommand{\L}
  {{\cal L}}
\newcommand{\M}
  {{\cal M}}
\newcommand{\N}
  {{\cal N}}
\renewcommand{\O}
  {{\cal O}}
\renewcommand{\P}
  {{\cal P}}
\newcommand{\Q}
  {{\cal Q}}
\newcommand{\R}
  {{\cal R}}
\renewcommand{\S}
  {{\cal S}}
\newcommand{\T}
  {{\cal T}}
\newcommand{\U}
  {{\cal U}}
\newcommand{\V}
  {{\cal V}}
\newcommand{\W}
  {{\cal W}}
\newcommand{\X}
  {{\cal X}}
\newcommand{\Y}
  {{\cal Y}}
\newcommand{\Z}
  {{\cal Z}}
\newcommand{\uno}
  {\unity}
\newcommand{\lie}
  {\ell_\varepsilon}
\newcommand{\lt}[1]
  { \ell_{t_{#1}} }
\newcommand{\leps}[1]
  { \ell_{\varepsilon_{#1}} }
\newcommand{\KG}
{\mbox{ \tiny \begin{tabular}{|c|} \hline \\ \hline \end{tabular} } }
\newcommand{\sedici}
{\overline{16}}

\newcommand{\alp}{{\dot\alpha}}
\newcommand{\bet}{{\dot\beta}}
\newcommand{\gamm}{{\dot\gamma}}
\newcommand{\delt}{{\dot\delta}}
\newcommand{\epsilo}{{\dot\epsilon}}
\newcommand{\et}{{\dot\eta}}
\newcommand{\iot}{{\dot\iota}}
\newcommand{\kapp}{{\dot\kappa}}
\newcommand{\lambd}{{\dot\lambda}}
\newcommand{\muu}{{\dot\mu}}
\newcommand{\nuu}{{\dot\nu}}
\newcommand{\sigm}{{\dot\sigma}}
\newcommand{\ta}{{\dot\tau}}
\newcommand{\omeg}{{\dot\omega}}
\newcommand{\ep}{\varepsilon}
\newcommand{\lamb}{\tilde\lambda}

\newcommand{\ansatz}{{\it Ans\"atze }}


\newcommand{\PropI}
{
\Delta^{-1}_{\Lambda\Lambda_0}
}
\newcommand{\Prop}
{
\Delta_{\Lambda\Lambda_0}
}
\newcommand{\rif}[1]
  {(\ref{#1})}
\newcommand{\ebarra}
  {$\bar e\ $}
\newcommand{\limPoincare}
  {$\bar e\longrightarrow 0\ $}
\newcommand{\inbasso}[1]
  { _{{\cal #1}} }
\newcommand{\inalto}[1]
  { ^{{\cal #1}} }
\newcommand{\call}[1]
  {{\cal #1}}
\newcommand{\eps}[1]
  {\varepsilon_{a_1 \cdots a_D}}
\newcommand{\xx}[2]
  {\frac {1}{2} #1_{aa'}#2^{aa'}}
\newcommand{\xxx}[2]
  {\frac 12 #1_{abc}#2^{abc}}
\newcommand{\xxxx}[3]
  { \frac12#2_{a_1\cdots a_{#1}}#3^{a_1\cdots a_{#1}} }
\newcommand{\formula}[2]
  { \begin{equation} \label{#1} #2 \end{equation} }
\newcommand{\formule}[2]
  { \begin{align} \label{#1} #2 \end{align} }
\newcommand{\formulona}[2]
{
  \begin{equation} \label{#1}
  \begin{split}
    #2 
  \end{split}
  \end{equation}
}
\newcommand{\formulonaX}[1]
{
  \begin{equation} 
    \begin{split}
      #1
    \end{split}
  \end{equation}
}
\newcommand{\graffa}[2]
{
  \begin{equation} \label{#1}
    \left\{ \begin{array}{lll} #2 \end{array} \right.
  \end{equation}
}
\newcommand{\graffaetichettata}[3]
{
  \begin{equation} \label{#2}
    #1\ \left\{ \begin{array}{lll} #3 \end{array} \right.
  \end{equation}
}
\newcommand{\formulaX}[1]
{  \begin{equation} #1 \end{equation} 
}
\newcommand{\formuleX}[1]
{ 
  \begin{align} #1 \end{align} 
}
\newcommand{\graff}[1]
{
  $$  \left\{ \begin{array}{lll} #1 \end{array} \right. $$
}
\newcommand{\graffet}[2]
{
 $$  #1\ \left\{ \begin{array}{lll} #2 \end{array} \right. $$
}
\newcommand{\graffaX}[1]
{
  \begin{equation} 
    \left\{ \begin{array}{lll} #1 \end{array} \right.
  \end{equation}
}
\newcommand{\graffaetichettataX}[2]
{
  \begin{equation}
    #1\ \left\{ \begin{array}{lll} #2 \end{array} \right.
  \end{equation}
}
\newcommand{\tabella}[4]
{
    \label{#1}
    \begin{center}
      #2\\ $\ $ \\
      \begin{tabular}{#3}
        #4  
      \end{tabular}
    \end{center}
}
\newcommand{\tabarray}[4]
{
    \label{#1}
    \begin{center}
      #2 $$ \begin{array}{#3} #4 \end{array} $$
    \end{center}
}
\newcommand{\tabellaX}[3]
{
  \begin{table}
   \caption{#1}
    \begin{center}
      \begin{tabular}{#2}#3
      \end{tabular}
    \end{center}
  \end{table}
}
\newcommand{\tabarrayX}[4]
{
  \begin{table}
    \caption{#2}
    \label{#1}
    \begin{center}
       $$ \begin{array}{#3} #4 \end{array} $$
    \end{center}
  \end{table}
}
\newcommand{\detizero}[1]
{
\left.{d\over dt}#1 \right|_{t=0}
}
\newcommand{\prendi}[2]
{
\left. #1 \right|_{#2}
}
\newcommand{\dezero}[2]
{
\left.{\mbox{d} #1\over \mbox{d} #2} \right|_{#2=0}
}
\newcommand{\deltazer}[2]
{
\left.{\delta#1\over\delta#2}\right|_{\breve \mu=0}
}
\newcommand{\deltazero}[2]
{
\left.{\delta #1\over \delta #2} \right|_{#2=0}
}
\newcommand{\dev}
{ \deltazer Iv }
\newcommand{\depsi}[1]
{ \deltazer I{\bar\psi^{#1}} }
\newcommand{\dew}
{ \deltazer I\omega }
\newcommand{\deA}
{ \deltazer IA }
\newcommand{\deB}
{ \deltazer IB }
\newcommand{\deff}
{ \deltazer If }
\newcommand{\desude}[1]
{
{\partial \over \partial #1}
}
\newcommand{\dxx}[2]
{
{ \partial x^{#1}\over\partial x^{#2} }
}
\newcommand{\dd}[2]
{
\frac{\delta{#1}}{\delta{#2}}
}
\newcommand{\dede}[2]
{
{\partial{#1}\over\partial{#2}}
}
\newcommand{\dedi}[2]
{
{\mbox{d}{#1}\over \mbox{d}{#2}}
}
\newcommand{\dexx}[2]
{
{ \partial x^{#1}\over\partial x^{#2} }
}
\newcommand{\dexy}[1]
{
{ \partial x^{#1}\over\partial x^{#1'} }
}
\newcommand{\deyx}[1]
{
{ \partial x^{#1'}\over\partial x^{#1} }
}

	
\begin{titlepage}
\renewcommand{\thefootnote}{\fnsymbol{footnote}}
\begin{flushright}
     LPTHE 98-10\\
     October 1998 \\
\end{flushright}
\par \vskip 10mm
\begin{center}
{\Large \bf
Gauge Consistent Wilson Renormalization Group II: Non-Abelian case}
\end{center}
\par \vskip 2mm
\begin{center}
{\large M.\ Simionato}\footnote{
E-mail: micheles@lpthe.jussieu.fr
}\\
\vskip 5 mm
{\small \it
LPTHE, Universit\'e Pierre et Marie Curie (Paris VI) et Denis Diderot 
(Paris VII), Tour 16, $1^{er}$ 
\'etage, 4, Place Jussieu, 75252 Paris, Cedex 05, 
France  and Istituto Nazionale di Fisica Nucleare, Frascati, Italy }
\end{center}
\par \vskip 2mm
\begin{center} {\large \bf Abstract} 
\end{center}
{\small \begin{quote}
We give a Wilsonian formulation
of non-Abelian gauge theories explicitly consistent with axial gauge
Ward identities. 
The issues of unitarity and dependence on the quantization direction 
are carefully investigated. A Wilsonian computation of the one-loop
QCD beta function is performed.\\

\end{quote}}
\end{titlepage}

\section{Introduction}

In a recent paper \cite{paper.I} we succeeded to give a continuous
Wilsonian formulation of Abelian gauge theories explicitly consistent 
with gauge-invariance. 
The basic idea was of introducing the Wilsonian infrared
cutoff $\Lambda$, which separates the soft from the hard modes, 
as a mass term for the propagating fields. 
In this way the Wilson's Renormalization Group Equation (also called Exact
Renormalization Group Equation, ERGE) becomes consistent with
the Ward-Takahashi identities to all scales.
The price to pay is the need for an explicit regularization
and renormalization of the evolution equation, since with the mass cutoff
the ultraviolet momenta are not sufficiently suppressed. However this is a
little price to pay since, at least in perturbation theory, there exist
gauge-invariant techniques to manage the ultraviolet divergences (dimensional
regularization with minimal subtraction, higher derivatives approaches, etc.).

In this paper we introduce a suitable generalization 
working in the non-Abelian case. The key idea 
is to use an algebraic non-covariant gauge.
This is quite natural since in that kind of gauges we can recover as much as
possible the properties of the Abelian Ward identities and
we can implement the method developed in \cite{paper.I}.
There are various gauges belonging to the general class of 
algebraic non-covariant gauges \cite{Bassetto,Leibbrandt}: the
axial gauge, the planar gauge and the light-cone gauge. 
Of these the light-cone gauge is the most famous and the most solid from a
theoretical point of view. Nevertheless, in this paper, we will
restrict to the analysis of the pure axial gauge, which is 
technically the simplest to manage. Actually, it is known that
this choice has non-trivial infrared problems related to a
proper definition of the spurious singularities: in
particular the Wilson loop consistency tests fails with the 
Cauchy Principal Value (CPV) prescription
\cite{Soldati}. 
In that paper we will use a regularization of spurious divergences
different from the CPV prescription, nevertheless we still expect
a very delicate infrared limit. The study of that problem, which
eventually could require the switching to a more established gauge,
as the light-cone one, will be the subject of a forthcoming paper
\cite{paper.III}. Herein, for sake of space, we will restrict 
to the analysis of Ward identities and ultraviolet properties. 
A great amount of what we will discuss for the pure-axial gauge case 
can be extended to the other non-covariant gauges.

Non-covariant gauges have been implemented in the 
ERGE formalism in \cite{Litim} (for the axial gauge)
and \cite{Geiger} (for the light-cone gauge). However, in these references
a generic Wilsonian cutoff function has been employed.
As a consequence the standard Ward identities are broken. 
In other words, even if we start from a gauge-invariant
ultraviolet action, the evolution generates non-gauge-invariant 
structures (spurious couplings).
Actually these couplings are constrained by some 
modified Ward identities (mWI's) \cite{Litim,Ellw,Datta} which
becomes the usual ones only at the physical scale $\Lambda=0$.
Although these mWI's are perfectly understood at the formal level, 
they are quite difficult to study in practice. 
In particular, in order to perform a consistent computation, 
we should provide an initial (ultraviolet)
action consistent with the mWI's. From a perturbative point of view, this 
is not a problem since it is well known \cite{Datta,Becchi} that it is 
possible to choose the boundary conditions on the spurious couplings in 
terms of the physical couplings in such a way that the mWI's hold to 
all scales.
This property is a consequence of the Quantum Action Principle \cite{QAP} and
holds to any order in perturbation theory, provided that the theory is 
anomaly free.
However, this approach involves a highly non-trivial fine-tuning procedure 
which is extremely cumbersome beyond the one-loop
level\footnote{Even at one-loop the fine-tuning is not so simple; 
see \cite{BDM.YM}
for an analysis in the covariant Feynman gauge.}.
Moreover at the non-perturbative level no non-trivial truncations 
of the effective action consistent with the mWI's are known. 
The same problem also holds in covariant gauges
and it is even worse in this case, since the modified Slavnov-Taylor
identities (mST's) \cite{Ellw,Datta} are more cumbersome.

The solution of this technical difficulty motivates our work. In fact, we will
prove that, by using a specific choice of the infrared cutoff, 
the {\it usual} form of the Ward identities can be maintained to {\it all} 
scales. In other words, in our formulation the evolution equation is
gauge-invariant and  the renormalization group 
flow preserves the gauge-symmetry. Two different proofs and a practical check
of this issue will be provided in this paper.

Moreover, the question of unitarity will be examined in detail. \\
Clearly, in a generic 
Wilsonian procedures unitarity is spoiled due to the breaking of gauge 
symmetry. In our formalism instead unitarity can be maintained
for any $\Lambda$, even in the non-Abelian case. This fact will be
proved by explicitly showing the compatibility of the propagator with the 
Landau-Cutkosky rules. However, as it  should be expected, 
the theory is {\it not} physically consistent
at $\Lambda\neq0$, since there is an unavoidable dependence
of ``physical'' quantities on the quantization direction $n_\mu$. 
In other words there is 
an unphysical breaking of Lorentz-covariance. In order to control this 
problem (which is generic for
any Wilsonian formulation of the axial gauge) we introduce a
generalized BRST symmetry holding also at $\Lambda\neq0$. 
In this way a simple analysis of the $n_\mu-$dependence is possible in terms of
generalized Slavnov-Taylor identities. 
In particular we show formally (i.e. modulo problems with 
infrared divergences, which will be addressed in \cite{paper.III}) 
that when the 
infrared cutoff is removed the Lorentz-covariance of the physical theory 
is recovered. We stress that in general the control of the unphysical 
$n_\mu-$dependence cannot be neglected since in numerical analysis 
the $\Lambda\to0$ limit is never reached.

Finally, we check the practical reliability of the formulation with
some perturbative computation: in particular we recover,
with a Wilsonian non-standard computation, the usual universal value of 
the one-loop QCD beta function. 

The plan of the paper is the following:
section 2 contains our notations and conventions;
section 3 is a simple introduction to the 
axial gauge in presence of an explicit
mass term; in section 4 we briefly discuss the 
Wilsonian (perturbative) renormalization of the theory. 
In section 5 we check the Ward identities
at one-loop with elementary methods: 
in particular we study the transversality property of
the gluon propagator. In section 6 and 7 we give the general 
proof of gauge invariance by following two different approaches.
Section 8 faces the problems of unitarity and gauge-dependence at the
formal level, by using BRST techniques.
In section 9, as a consistency check of the formalism,
we compute the one-loop QCD beta function with Wilsonian techniques.
Section 10 contains our conclusions.
Three appendices concerning technical points close the paper. 
Appendix A contains some comments about the renormalizability property
in the axial gauge framework from a Wilsonian point of view. In appendix B 
the rigorous deduction of the modified
Ward identities is reviewed and the relation with the Quantum Action Principle
and the fine tuning problem is clarified.
Appendix C contains some useful trick
to make perturbative computations in our formalism 
in an efficient way.

\section{Notations and conventions}

For future reference and for commodity of the reader, we begin by collecting
our conventions and some useful formula.\\
We work in QCD with $N_C$ colors and $N_f$ flavors.
The gauge fields are 
denoted by $A^a_\mu(x)$;
the matter fields (quarks and antiquarks) with $\psi^i(x)$ and 
$\bar\psi_i(x)$, where $i$ is 
the flavor index; spinor and color indices are understood. The
generators of the group in the fundamental representation are denoted by
$T_a$ and are taken hermitian $T_a=T_a^+$; 
the generators in the adjoint representation are denoted by $\tau_a$ and 
corresponds to the structure constants  $(\tau_a)_{bc}=f_{abc}$. 
We use the condensed notations $X\cdot Y= 
\delta_{ab}X^a Y^b$ and  $(X\times Y)^a =f^a_{bc}X^b Y^c$ 
for the inner and the outer product in the adjoint representation. 
The covariant derivative $D_\mu$ 
acts as $D_\mu X=\partial_\mu X+g A_\mu\times X$ and
$D_\mu\psi=\partial_\mu\psi-i gA_\mu^a T_a\psi$; $g$ is the
coupling constant.
The gauge transformations
are generated by the functional operator
\formula{def.W}
{\W_f=-\int d^4x f(x)\cdot\left[D_\mu\dd{}{A_\mu}+ig \bar\psi T\dd{}{\bar\psi}-
igT\psi\dd{}{\psi}\right]
}
and on the fundamental fields we have the relations
\formula{gt}
{\W_f A_\mu= D_\mu f,\quad \W_f \psi=igf\cdot T\psi,\quad 
\W_f \bar\psi=-igf\cdot \bar\psi T.} 
The field strength tensor is defined
as $F_{\mu\nu}=\partial_\mu A_\nu-\partial_\nu A_\mu+ gA_\mu\times A_\nu$
and the gauge-invariant classical action is
\formula{cl.action}
{S_{CL}(A,\psi,\bar\psi)=
S_{CL,gauge}(A)+S_{CL,matter}(A,\psi,\bar\psi)}
with
\formulaX
{S_{CL,gauge}(A)=-\frac14\int d^4x\ F_{\mu\nu}\cdot F^{\mu\nu},}
\formulaX
{S_{CL,matter}(A,\psi,\bar\psi)=\int d^4x\ \bar\psi_i (i\slash D-m^i_j)\psi^j.}
Here $m^i_j$ denotes the (diagonal) quark mass matrix.
We will use the tensors
\formula{t.l}
{t_{\mu\nu}(p)=g_{\mu\nu}-p_\mu p_\nu/p^2,\quad \ell_{\mu\nu}=p_\mu p_\nu/p^2,
}
\formula{V...}
{V_{\mu\nu\rho}(p,q,r)=(q-r)_\mu g_{\nu\rho}+(p-q)_\rho g_{\mu\nu}+(r-p)_\nu 
g_{\rho\mu},
}
\formulona{t4}
{t_{\mu\nu\rho\sigma}^{abcd}&=(f^{acbd}-f^{adcb})g_{\mu\nu}g_{\rho\sigma}+
(f^{abcd}-f^{adbc})g_{\mu\rho}g_{\nu\sigma}+\\
&\quad(f^{acdb}-f^{abcd})g_{\mu\sigma}g_{\nu\rho}, \quad
f^{abcd}\equiv f^{ab}_e f^{ecd}.
}
The following classical Ward identities holds, as a consequence of
gauge-invariance:
\formula{W.cl.1}
{p^\mu t_{\mu\nu}(p)=0}
\formula{W.cl.2}
{p_1^\mu V_{\mu\nu\rho}(p_1,p_2,p_3)=p_3^2 t_{\nu\rho}(p_3)-p_2^2 
t_{\nu\rho}(p_2)}
\formulona{W.cl.3}
{p_1^\alpha t^{abcd}_{\alpha\beta\gamma\delta}=&-f^{abcd}V_{\beta\gamma\delta}
(p_1+p_2,p_3,p_4)-f^{acdb}V_{\gamma\delta\beta}(p_1+p_3,p_4,p_2)\\
&-f^{adbc}V_{\delta\beta\gamma}(p_1+p_4,p_2,p_3).
}
In \rif{W.cl.2} and \rif{W.cl.3} the vanishing of the sum 
$\sum_i p_i=0$ is understood.
These identities will be used in the study of the one-loop 
quantum Ward identities.
In the text we will use the deWitt condensed notation, by denoting with
$$\Phi^A=(A_\mu^a(p),\psi_i(p),\bar\psi^i(p))$$ 
all the fields of theory (when required in BRST considerations, we will also
collect in $\Phi^A$ the ghosts and the auxiliary fields). 
The 1PI Green functions are generally denoted by 
\formula{vertices}
{\Gamma_{A_1\dots A_n}=
\left.\dd\Gamma{\Phi^{A_1}\dots \delta\Phi^{A_n}}\right|_{\Phi=0}.
}
For instance the two-point function $\Gamma_{AB}$ both corresponds to the
gluon propagator $(2\pi)^4\delta(p+q)\Gamma_{\mu\nu}(p)$ and to the quark
propagator $(2\pi)^4\delta(p+q)\Gamma_{ij}(p)$. 
Sums and integrals over repeated indices 
are understood and the supertrace notation is used,
$\STr X=(-)^A X^A_A$, where $(-)^A$ has the value $+1$ for bosonic fields
and $-1$ for fermionic fields.

Some useful abbreviations on integrals are
\formula{int.abbr}
{\int_x=\int d^4x,\quad\int_q=\int\frac{d^4q}{(2\pi)^4},\quad
\int_{q_3}=\int\frac{dq_3}{2\pi},\quad\int_{\bar q}=
\int\frac{d^3\bar q}{(2\pi)^3},
}
with $\bar q=(q^0,q^1,q^2)$.
For Euclidean vectors we shall use the notations 
\formula{Euclidean}
{p_E=(i p_0,\vec p),\quad q_E=(i q_0,\vec q),\quad p_E q_E
\equiv\delta_{\mu\nu}\ p_E^\mu\ q_E^\nu=-g_{\mu\nu}p^\mu q^\nu=-p q.}
If not otherwise specified, all the quantities
should be intended in the Minkowski space.

Equipped with these conventions, we can begin our analysis with some
simple consideration on the gauge-fixing procedure from a Wilsonian
perspective.

\section{Remarks on the gauge-fixing procedure}

In order to define a perturbative
quantum field theory from a classical gauge invariant field theory one 
is forced to break gauge invariance. 
This breaking is required to define an invertible propagator 
for the gauge field and is technically done by adding to the classical 
(free) action 
\formula{S.CL}
{S_{CL}(A)=\int_x-\frac14F_{\mu\nu}F^{\mu\nu}=
-\int_p\frac12A_\mu(-p)t^{\mu\nu}(p)p^2A_\nu(p)} 
a gauge fixing term.
Typically, this term is quadratic in the gauge fields i.e. only affects the
propagators and not the vertices of the theory. 
Therefore the gauge fixed action has the general form
\formula{GF}
{\Gamma^{(0)}_{gauge}(A)=S_{CL}(A)+S_{GF}(A),\quad S_{GF}(A)=
\frac12\int_p A_\mu(p) Q^{\mu\nu}(p) A_\nu(p)
}
where $Q_{\mu\nu}(p)=Q_T(p) t_{\mu\nu}(p)+Q_L(p) 
\ell_{\mu\nu}(p)$ 
is a suitable symmetric tensor, to be taken
such as the propagator of the gauge fields
\formula{prop}
{-D_{\mu\nu}(p)=\left(p^2t_{\mu\nu}(p)-
Q_{\mu\nu}(p)\right)^{-1}=
\frac{t_{\mu\nu}(p)}{p^2-Q_T(p)+i\ep}+\frac{\ell_{\mu\nu}(p)}
{-Q_L(p)+i\ep}
} 
satisfies regularity properties both in the ultraviolet and in the
infrared.
The more general gauge fixing
quadratic in the fields consistent with Lorentz symmetry and the
power counting criterium is given by the formula \cite{paper.I}
\formula{Q.QED}
{Q_{\Lambda,\mu\nu}(p)=
-\frac1\xi p_\mu p_\nu+\Lambda^2g_{\mu\nu},
}
where $\xi$ is a generic\footnote{The only requirement is 
$\xi<\infty$ in order the propagator satisfies the power counting criterium
\rif{power.c}. In the limit $\xi\to\infty$ one recovers the Proca theory 
\cite{ZJ.book}.} 
dimensionless parameter which is expected do not affect the physics
whereas the scale $\Lambda$ is interpreted as a
Wilsonian cutoff distinguishing between soft ($p_E^2<<\Lambda^2$) and hard
($p_E^2>>\Lambda^2$) modes. With this choice the Euclidean propagator
becomes regular both in the infrared
\formula{infrared}
{D_{\Lambda,\mu\nu}(p_E)\sim 1/\Lambda^2,\quad p_E^2<<\Lambda^2,
}
and in the ultraviolet
\formula{power.c}
{D_{\Lambda,\mu\nu}(p_E)\sim 1/p_E^2,\quad p_E^2>>\Lambda^2,
} 
and in particular satisfies the power counting criterium
which is essential in the renormalizability proof.

In the Abelian case this simple procedure is enough to construct a
consistent quantum field theory in the framework of the ERGE \cite{paper.I}.
The reason of this success is clear: even if gauge
invariance is formally broken, nevertheless the theory still preserve
all the good properties of the gauge symmetry since the tree-level
Ward-Takahashi identities are only linearly broken,
\formula{Ward.0.QED}
{\W_f\Gamma^{(0)}(A,\xi;\Lambda)=\int_x f(x)\left
(\frac\square\xi+\Lambda^2\right)\partial_\mu A^\mu.
}
This exceptional property guarantees that identity \rif{Ward.0.QED}
can be lifted to the quantum level, provided that we use 
a consistent renormalization procedure~\footnote{Needless to say, 
such a procedure does not exist in many interesting cases, in particular in
chiral models.}. Essentially, in \cite{paper.I} we have
proved that the  Wilson's Renormalization Group Equation
provides such a consistent procedure.
Therefore perturbative corrections to the tree level action are gauge invariant
to all orders,
\formulaX
{\W_f\Gamma^{(\ell)}(A,\xi;\Lambda)=0,\quad\forall\ell\geq1.
}
As it is well known  \cite{ZJ.book} this fact assures the 
unitarity of the Abelian theory to all scales $\Lambda$.
However, in the non-Abelian case, this procedure does not work, since the 
breaking term is quadratic in the fields:
\formulaX
{\W_f\S_{GF}=\frac12\int_x A_\mu\cdot Q_\Lambda^{\mu\nu}D_\nu f+
D_\mu f \cdot Q_\Lambda^{\mu\nu} A_\nu.
}
For this reason the gauge symmetry cannot be lifted to
the quantum level and the quantum corrections are {\it not} gauge-invariant.
The standard way to solve this problem is of introducing the ghost fields
$C^a(x),\ \bar C^a(x)$ and the Nakanishi-Lautrup auxiliary fields $B^a(x)$,
substituting the gauge symmetry with the BRST symmetry \cite{BRST}
\formula{BRST.cov}
{s A_\mu= D_\mu C,\quad s C=-\frac12 g C\times C,\quad s\bar C=B,\quad sB=0.}
In this way we can take
\formula{S.BRST}
{S_{BRST}=S_{CL}+s\left(\bar C\cdot\partial_\mu A^\mu+\frac12\xi\bar C\cdot B
\right)
}
as a suitable tree level action from which a renormalization program
consistent with the symmetry can start. However,
there is a price to pay: the BRST symmetry is {\it non-linear} in
quantum fields and it can be lift to the quantum level only in the form of
Slavnov-Taylor identities which renormalization is technically more 
complicate. Moreover a detailed study of graphs involving ghost particles
is required in doing perturbative computations.
Actually, as it is well known, one can avoid these technical complications
if one relaxes the requirement of covariance and works in a particular
class of non-covariant gauges (actually this was the original motivation in
studying axial gauges, also called ghost-free gauges or physical gauges
\cite{Kummer}). 

Here we are interested in the pure axial gauge which has been extensively
used both in theoretical and in phenomenological literature
(for a review see \cite{Bassetto,Leibbrandt}).
However, since we have in mind a Wilsonian interpretation, our analysis
will be non-standard, based on a massive version of the
usual axial gauge-fixing. For definiteness we will take as gauge fixing term
\formula{GF.axial}
{S_{GF}(A,\xi_2;\Lambda)=
\int_x\frac1{2\xi_2 n^2}n^\mu A_\mu\cdot n^\nu A_\nu+
\frac12\Lambda^2A_\mu\cdot A^\mu,
}
where $n^\mu$ is a space-like vector, $n^2<0$. 
In particular the Arnowitt-Ficker
choice $n^\mu=(0,0,0,1)$ will be considered in the beta function computation
in section 9. In terms of the cutoff
function $Q_{\Lambda,\mu\nu}$ the axial gauge fixing corresponds to the choice
\formula{Q.axial}
{Q_{\Lambda,\mu\nu}=g_{\mu\nu}\Lambda^2+\frac1{\xi_2}
\frac{n_\mu n_\nu}{n^2}.
}
Notice that in \rif{GF.axial} we have introduced a non-renormalizable
auxiliary parameter $\xi_2$ of mass dimension $-2$. However in this paper 
we will only consider the limit $\xi_2\to0$ (pure axial gauge\footnote{
One can easily convince himself that $\xi_2=0$ is a fixed point of the
ERGE, and actually this property holds  for any choice of the cutoff
function \cite{Litim}. Therefore $\xi_2=0$ is preserved by the evolution.}).
In this case the propagator has the form
\formulona{D.axial}
{-D_{\Lambda,\mu\nu}(p)
\buildrel{\xi_2 \to 0}\over=\frac1{ p^2-\Lambda^2+i\ep}g_{\mu\nu}-
\frac{p\cdot n(n_\mu p_\nu+n_\nu
p_\mu)}{(p^2-\Lambda^2+i\ep)((p\cdot n)^2-n^2\Lambda^2)}+\\
\frac{p_\mu p_\nu\ n^2}{(p^2-\Lambda^2+i\ep)((p\cdot n)^2-n^2\Lambda^2)}+
\frac{\Lambda^2 n_\mu n_\nu}
{(p^2-\Lambda^2+i\ep)((p\cdot n)^2-n^2\Lambda^2)}.
}
The limit $\xi_2\to0$ 
has particular properties since  $D_{\Lambda,\mu\nu}(p)$  
becomes transverse
\formula{trans}
{n^\mu D_{\Lambda,\mu\nu}(p)\buildrel{\xi_2\to 0}\over=0,\quad\forall
\ \Lambda
}
and therefore is not invertible (for any $\Lambda$).
This is the same phenomenon which happens in the covariant Landau gauge.
The na\"\i ve zero mass limit of \rif{D.axial} gives the usual form of the
axial gauge propagator \cite{Kummer}: however this limit is quite
delicate since there are spurious singularities at $n^\mu p_\mu=0$.
Such infrared singularities are automatically regularized at $\Lambda\neq0$.
This is a typical feature of the Wilsonian approach, also 
noticed in \cite{Litim}. These authors in fact computed the general
form of the propagator with a generic cutoff $R_{\Lambda,\mu\nu}(p)$ of which
\rif{trans} is a particular case. 
However, this particular case is
exceptionally important since in this case 
the gauge-fixing term breaks gauge invariance only in a linear way
\footnote
{In the deduction of \rif{fondamentale}
the identities $n_\mu A^\mu\cdot n_\nu A^\nu\times f=
n_\mu n_\nu A^\mu\times A^\nu\cdot f=0$ and $A_\mu \times A^\mu\cdot f=0$.
should be used.}
\formula{fondamentale}
{\W_f S_{GF}=\int_x\frac1{\xi_2 n^2}n_\mu A^\mu\cdot n_\nu\partial^\nu f+
\Lambda^2A^\mu\cdot\partial^\nu f=\int_x A^\mu\cdot 
Q_{\Lambda,\mu\nu}\partial^\nu f.
}
Therefore in this case the situation is QED-like and 
the classical gauge symmetry can be lifted to the quantum level in 
a consistent way. 

The same result holds when the matter coupling
is considered, provided that one introduces the Wilsonian infrared
cutoff as a mass term for the matter fields, as has been done in 
\cite{paper.I}, by taking
\formula{matter}
{\Gamma^{(0)}_{matter}(A,\bar\psi,\psi;\Lambda)=
\int_x\bar\psi_i i\slash D\psi^i-
\bar\psi_i(m^i_j+i\Lambda\gamma_5\delta^i_j)\psi^j.
}
In the following we will also use the general notation\footnote{A more 
explicit notation is $\frac12\tilde\Phi Q_\Lambda\Phi=\frac12(-)^A\Phi^A 
Q_{\Lambda,AB}\Phi^B=\frac12Q_{\Lambda,AB}\Phi^B\Phi^A$. The tilde
denotes the transposition of the generalized indices with suitable 
minus signs for fermionic entries.}
\formula{tree.level}
{\Gamma^{(0)}(\Phi;\Lambda)=S_{CL}(\Phi)+\frac12\tilde\Phi Q_\Lambda\Phi,
}
with
\formula{def.Q}
{\frac12\tilde\Phi Q_\Lambda\Phi\equiv
\frac12\int_p A_\mu(-p)Q^{\mu\nu}_\Lambda(p)A_\nu(p)-
\int_p\bar\psi_j(p)i\Lambda\gamma_5\psi^j(p),
}
since a geometric interpretation becomes more clear: 
the (graded) symmetric invertible tensor $Q_{\Lambda,AB}$
can be seen as a metric in the field space. Thus, 
there is an explicit invariance of the gauge-fixing term under isometries, 
i.e. linear transformation $\Phi^A\to U^A_B\Phi^B$ such as
$\tilde U_C^A Q_{\Lambda,AB} U^B_D=Q_{\Lambda,CD}$.
This interpretation will be useful to
give a clear proof of gauge-invariance in section 6.
 
\section{Wilsonian perturbative renormalization}

Having defined the tree level theory, i.e. the functional 
$\Gamma^{(0)}(\Phi;\Lambda)$, one can compute the quantum correction 
$\Gamma^{(\ell)}(\Phi;\Lambda)$ to any order in perturbation theory
by introducing a consistent renormalization procedure.
In this paper we shall adopt the renormalization
procedure based on the ERGE introduced
in \cite{paper.I} (see also
\cite{Wetterich1,BDM,Morris1}).
By referring to \cite{paper.I} for a complete discussion, here
we simply write down the explicit form of the equation for the proper vertices,
which reads
\formula{PRGE}
{\dot\Pi_{A_1\dots A_n}=\hbar I_{A_1\dots A_n}=-\hbar\left[\frac i2(-)^A
(\Gamma_2^{-1}\dot Q_\Lambda\Gamma_2^{-1})^{BA}
\bar\Gamma_{AA_1\dots A_nB}\right]_{reg}.
}
We remind the notation:
\begin{itemize}
\item $\Pi(\Phi;\Lambda)$ is the gauge-invariant effective action
\formula{def.Pi}
{\Pi(\Phi;\Lambda)\equiv\Gamma(\Phi;\Lambda)-\frac12\tilde\Phi Q_\Lambda\Phi
}
which satisfies the non-Abelian Ward-Takahashi identities
\formula{WTNA}
{\W_f\Pi=\int_x
D_\mu f\cdot \dd\Pi{A_\mu}=-\int_xf\cdot\left(\partial_\mu\dd\Pi{A_\mu}
+g A_\mu\times \dd\Pi{A_\mu}\right)=0}
as we will prove.
\item The auxiliary vertices $\bar\Gamma_{AA_1\dots A_nB}$ are recursively
obtained from the usual vertices via the formula \cite{BDM} 
\formula{bar.Gamma.n.gen}
{\bar\Gamma_{AA_1\ldots A_nB}=\Gamma_{AA_1\ldots A_nB}-\sum_{k=1}^{n-1}
\Gamma_{AA_1\dots A_k C}(\Gamma_2^{-1})^{CD}\bar\Gamma_{DA_{k+1}\ldots A_nB}.
}
as shown in figure 1.

\item The subscript $reg$ in \rif{PRGE} denotes some
ultraviolet regularization needed to properly define the loop integrals 
in the evolution equation. In fact, as we explained in detail in 
\cite{paper.I}, in four dimensions the two-point functions $\dot\Pi_{AB}$ are 
logarithmically divergent and must be regularized and renormalized. 
In the following we
shall consider both the effects of a gauge-consistent and of a 
gauge-inconsistent regularization.
\end{itemize}

A graphical representation of the evolution equation is reported in figure~2.

\begin{figure}[t]                              
\setlength{\unitlength}{1.25 ex}
\begin{center}
\begin{picture}(44,8)

 \put(3,5){\framebox(4,2)}
 \puntino{1,6}{A}\put(1,6){\line(1,0){2}}
 \put(9,6){\line(-1,0){2}}\puntino{9,6}{B}
 \put(3,2){\line(1,3){1}}\puntino{3,2}{A_1}
 \puntino{7,2}{A_n}\put(7,2){\line(-1,3){1}}

 \scrivi{11,6}{=} \put(17,6){\oval(4,2)}
 \puntino{13,6}{A}\put(13,6){\line(1,0){2}}
 \put(21,6){\line(-1,0){2}}\puntino{21,6}{B}
 \put(15,2){\line(1,3){1}}\puntino{15,2}{A_1}
 \puntino{19,2}{A_n}\put(19,2){\line(-1,3){1}}

 \scrivi{23.5,6}{-} \put(30,6){\oval(4,2)}
 \puntino{26,6}{A}\put(26,6){\line(1,0){2}}
 \put(34,6){\line(-1,0){2}}\puntino{34,6}{CD}
 \put(28,2){\line(1,3){1}}\puntino{28,2}{A_{i_1}}
 \puntino{32,2}{A_{i_k}}\put(32,2){\line(-1,3){1}}

 \put(36,5){\framebox(4,2)}
 \put(42,6){\line(-1,0){2}}\puntino{42,6}{B}
 \put(34,6){\line(1,0){2}}\puntino{34,6}{CD}
 \put(36,2){\line(1,3){1}}\puntino{36,2}{A_{i_{k+1}}}
 \puntino{40,2}{A_n}\put(40,2){\line(-1,3){1}}
 
\end{picture} \caption{{\small Recursive expansion of the
$\bar\Gamma_{AA_1\dots A_nB}$ vertices, denoted by the boxes. 
The black dots denote the full propagators and the ovals the full vertices. 
A sum over inequivalent permutations of external lines is understood.}}
\end{center}
\end{figure}
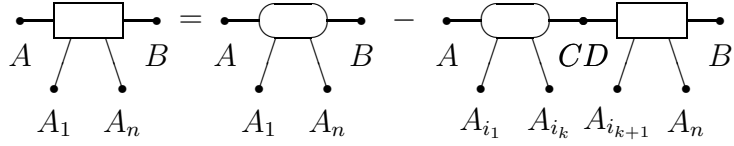

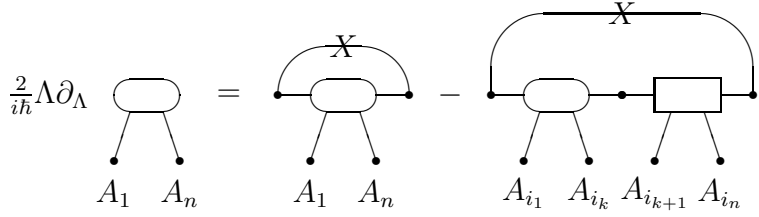
\begin{figure}[t]                              
\setlength{\unitlength}{1.2 ex}
\begin{center}
\begin{picture}(40,12)

 \scrivi{-1,6}{\frac 2{i\hbar}\LdL}
 \put(5,6){\oval(4,2)}\scrivi{5,6}{}
 \put(3,2){\line(1,3){1}}\puntino{3,2}{A_1}
 \puntino{7,2}{A_n}\put(7,2){\line(-1,3){1}}

 \put(17,6){\oval(8,6)[t]} \scrivi{17,9}{X}
 \scrivi{10,6}{=} \put(17,6){\oval(4,2)}\scrivi{17,6}{}
 \puntino{13,6}{}\put(13,6){\line(1,0){2}}
 \put(21,6){\line(-1,0){2}}\puntino{21,6}{}
 \put(15,2){\line(1,3){1}}\puntino{15,2}{A_1}
 \puntino{19,2}{A_n}\put(19,2){\line(-1,3){1}}

 \put(34,6){\oval(16,10)[t]} \scrivi{34,11}{X}
 \scrivi{23.5,6}{-} \put(30,6){\oval(4,2)}\scrivi{30,6}{}
 \puntino{26,6}{}\put(26,6){\line(1,0){2}}
 \put(34,6){\line(-1,0){2}}\puntino{34,6}{}
 \put(28,2){\line(1,3){1}}\puntino{28,2}{A_{i_1}}
 \puntino{32,2}{A_{i_k}}\put(32,2){\line(-1,3){1}}
 \put(36,5){\framebox(4,2)}
 \put(42,6){\line(-1,0){2}}\puntino{42,6}{}
 \put(34,6){\line(1,0){2}}\puntino{34,6}{}
 \put(36,2){\line(1,3){1}}\puntino{36,2}{A_{i_{k+1}}}
 \puntino{40,2}{A_{i_n}}\put(40,2){\line(-1,3){1}}
\end{picture} 
\end{center} \caption{{\small
Diagrammatic version of the exact 
evolution equation. Here $X=\dot Q_\Lambda$.}}
\end{figure}

Notice that in \rif{PRGE} we have explicited the factor $\hbar$
in order to clarify how the ERGE acts as an iterative
renormalization  procedure. In fact, as fully reviewed in \cite{paper.I}
(see also \cite{Polchinski} and \cite{BDM}) once boundary conditions
on relevant couplings (i.e. power counting renormalizable couplings 
in a more common terminology)
are fixed, one can solve iteratively the ERGE
to all orders in perturbation theory by recovering in practice the
BPHZ subtracted proper vertices. In the case at hand, since Lorentz
covariance is broken, the analysis of relevant couplings differs
from the covariant one and the more general gauge-invariant relevant
functional has the form
\formula{general.form}
{\Pi_{rel}(\Phi;\Lambda)=\int_x-\frac14 Z_A(\Lambda) 
F_{\mu\nu}\cdot F^{\mu\nu}-
\frac12 h(\Lambda) \frac{n^\mu n_\sigma}{n^2}F_{\mu\rho}\cdot F^{\rho\sigma}.
}
The relevant parameters $Z_A(\Lambda)$ and $h(\Lambda)$
can be obtained from the
two-point function which has the form\footnote{
Notice that both $t_{\mu\nu}(p)$ and $T_{\mu\nu}(p,n)$ are transverse
as a consequence of the Ward identity $p^\mu\Pi_{\mu\nu}(p,n;\Lambda)=0$.
}
\formula{self-energy}
{\Pi_{\mu\nu,rel}(p,n;\Lambda)=Z_A(\Lambda) p^2t_{\mu\nu}(p)+ h(\Lambda) 
p^2T_{\mu\nu}(p,n)}
with
\formula{TT}
{T_{\mu\nu}(p,n)=\frac{(n p)^2}{n^2p^2}g_{\mu\nu}-\frac{np}{n^2p^2}
(pn+np)_{\mu\nu}+\frac{n_\mu n_\nu}{n^2}.}

Here we adopt zero-momentum prescriptions and we fix the relevant couplings  
at some non-zero infrared renormalization scale 
$\Lambda_R$:
\formula{prescr}
{Z_A(\Lambda_R)=1,\quad h(\Lambda_R)=0.} 

The generation of the new relevant coupling $h(\Lambda)$ is due to the 
Lorentz-covariance breaking. However, this coupling is {\it not} an independent
coupling and its evolution is constrained from a
non-linear functional identity describing the dependence on the quantization
direction $n_\mu$. Moreover at one-loop $h(\Lambda)$ is zero. 
All these issues will be examined in detail in section 7.
Here we are more interested on 
the function $Z_A(\Lambda)$ which is physically very important since 
it is related to the coupling constant\footnote{
Clearly the coupling $g(\Lambda_R)$ can be related to the $g(\mu)$ 
coupling of some momentum dependent prescription, $g(\mu)=
f(g(\Lambda_R),\mu/\Lambda_R)$ and the explicit form of $f$ can be computed
in perturbation theory.} 
$g(\Lambda)$ via the QED-like relation
\formula{QED.like}
{g(\Lambda)=\frac {g(\Lambda_R)}{Z_A^{1/2}(\Lambda)}.
} 
This relation is imposed from the gauge symmetry and is {\it exact} 
in our formalism whatever vertex 
is used to define $g(\Lambda)$ (at zero momentum the $\bar\psi A\psi$
vertex, the $AAA$ vertex and the $AAAA$ vertex give the same coupling,
since gauge-invariance is preserved).
 
The important point we want to stress is the following: since our formalism 
is Ward-identities-consistent, no spurious couplings
are generated and {\it no fine tuning} is required for them.

Having fixed the boundary conditions, in principle 
one can solve the evolution equation to
all orders in perturbation theory 
in terms of integrals over the scale $\Lambda$
since the renormalizability property guarantees that they 
are all well defined. This point is fully treated in \cite{paper.I} and 
can be lifted to the present
case without problems (we refer to appendix A for some additional comment). 
Instead, it
seems worthwhile of explaining how the  Renormalization
Group Equation works in practice, by
explicitly solving it at one-loop and by checking the
consistence with the Ward-Takahashi identities.
\begin{figure}[t]  
  \begin{center}
  \mbox{\epsfig{file=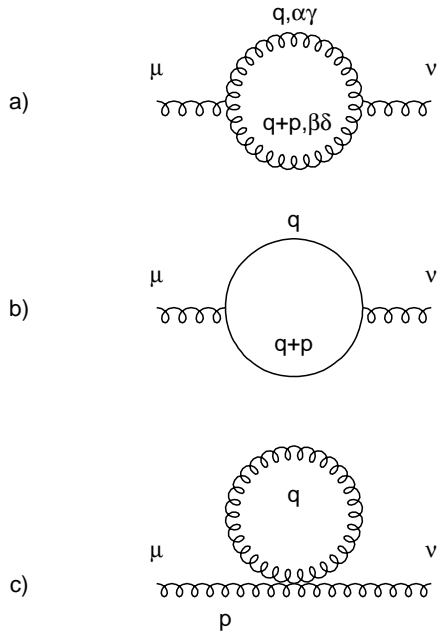,height=9cm}}
  \end{center}
\caption{{\small 
One-loop Feynman diagrams contributing to the evolution of the gluon
propagator. Color indices are understood.}}
\end{figure}
In particular we will study the evolution equation for the one-loop gluon
propagator, which has the general form
\formula{1.loop.gluon}
{\dot\Pi^{(1)}_{\mu\nu,ab}(p,n;\Lambda)=\LdL F_{\mu\nu,ab}^{reg}(p,n;
\Lambda)
}
where the function $F_{\mu\nu,ab}^{reg}(p,n;\Lambda)$ denotes the sum of 
the usual regularized one-loop Feynman diagrams (figure 3). 
For instance the contribution of figure 3a is given by
\formulonaX
{F^{(a)reg}_{\mu\nu,ab}(p,n;\Lambda)&=\frac{ig^2}2\delta_{ab}
\int_q[V^{\mu\alpha\beta}(  p,  q,-  q-  p)\\
&\quad\cdot D_{\Lambda,\beta\delta}(  q+  p)V^{\nu\delta\gamma}
(-  p,  q+  p,-  q)D_{\Lambda,\gamma\alpha}(  q) ]_{reg}.}
A completely
analogous term comes from the coupling with the matter fields (figure 3b)
where the gluon propagators are replaced by the quark propagators 
and a multiplicative factor of -2 is added.
Finally there is a contribution from the tadpole graph (figure 3c). 

A comment about the ultraviolet regularization is in order
here. In principle, due to the renormalizability property,
all ultraviolet renormalization are equivalent for the low energy 
predictions, however in practice some choice is simpler than others. 
In the following two sections 
we first discuss the case of a gauge-inconsistent 
regularization both with elementary diagrammatic methods
and with general theoretical methods. The advantages of a gauge-consistent
ultraviolet regularization, as for instance dimensional regularization or
an higher derivative regularization, are explained in section 7.\footnote{
Notice that with generic cutoff functions, there is no advantage in
using a gauge-consistent ultraviolet regularization since the cutoff
$\Lambda$ breaks the symmetry in the {\it infrared}: it is only with our 
specific choice that the analysis in section 7 is meaningful.}

\section{One-loop check of Ward identities} 

As starting point, let us consider the transversality property
of the gluon self-energy
\formula{transv.1}
{p^\mu\Pi_{\mu\nu}^{(1)}(p,n;\Lambda)=0.
}
If we fix the boundary conditions at the ultraviolet scale $\Lambda_0$
the self-energy is given by the evolution equation \rif{1.loop.gluon} as
\formula{Pi.1}
{\Pi_{\mu\nu}^{(1)}(p,n;\Lambda)=\Pi^{(1)}_{\mu\nu}(p,n;\Lambda_0)+
F_{\mu\nu}^{reg}(p,n;\Lambda)-F^{reg}_{\mu\nu}(p,n;\Lambda_0).
}
Now we have to specify the intermediate regularization.
For definiteness we regularize the ultraviolet behavior of the
(free) propagator by replacing
\formula{UV.reg}
{D_{\Lambda,\mu\nu}(q)\to D^{\Lambda_0}_{\Lambda,\mu\nu}(q)= 
\bar K_{0\Lambda_0}(q)D_{\Lambda,\mu\nu}(q),
}
where $\bar K_{0\Lambda_0}(q)$ is a suitable ultraviolet cutoff function,
for instance
\formula{K.UV}
{\bar K_{0\Lambda_0}(q)=e^{-q^2/\Lambda_0^2}.
}
Consider first the contribution from graph 3a. 
The proof of the transversality
property \rif{transv.1} is based on the tree level vertex identity
\formulona{tree.level.Ward}
{p^\mu V_{\mu\nu\rho}(p,q,-q-p)=D_{\Lambda,\nu\rho}^{-1}(q)-
D_{\Lambda,\nu\rho}^{-1}(q+p)
}
which has the same form of the classical identity \rif{W.cl.2}, thanks to
our good choice of the cutoff function $Q_{\Lambda,\mu\nu}$.
By using \rif{tree.level.Ward} we obtain
for the transverse part of the self-energy a contribution of kind
\formula{tras.a}
{p^\mu\Pi_{\mu\nu}^{(1)(a)}(p,n;\Lambda)=
p^\mu F^{(a)reg}_{\mu\nu}(p,n;\Lambda)-p^\mu F^{(a)reg}_{\mu\nu}
(p,n;\Lambda_0)+b.c.
}
where
\formulonaX
{p^\mu F^{(a)reg}_{\mu\nu}(p,n;\Lambda)&=
\frac{i g^2}2\int_q -D^{\gamma\delta}_\Lambda(q)
V_{\gamma\nu\delta}(-q,-p,q+p)\\ &\quad\cdot \bar K_{0\Lambda_0}(q)
\left[\bar K_{0\Lambda_0}(q+p)-\bar K_{0\Lambda_0}(q-p)\right]
}
and $b.c.$ denotes the boundary conditions on the transverse part
of the self energy at the scale $\Lambda_0$.
For dimensional and covariance reasons we can write
\formulona{dot.Delta.II}
{&p^\mu F^{(a)reg}_{\mu\nu}(p,n;\Lambda)-p^\mu F^{(a)reg}_{\mu\nu}
(p,n;\Lambda_0)=\\
&c_1^{(a)} p^2\ p n\ n_\nu+ c_2^{(a)} p^2\ p_\nu+c_3^{(a)}\Lambda^2p n\ n_\nu
+c_4^{(a)}\Lambda^2 p_\nu
+ O(\Lambda^2/\Lambda_0^2,p^2/\Lambda_0^2)
}
where the precise value of the numerical coefficients $c_i^{(a)}$ 
depends on the choice of the ultraviolet cutoff function 
$\bar K_{0\Lambda_0}(q)$ but it is not important here. 
The same argument holds when the coupling with the matter fields is
considered (figure 3b), 
since the analogous identity\footnote{$S_{\alpha\beta}(p;\Lambda)$
denotes the tree-level quark propagator.
}
\formulona{Ward.spin}
{p^\mu \gamma_{\mu\alpha\beta}=S_{\Lambda,\alpha\beta}^{-1}(q)-
S_{\Lambda,\alpha\beta}^{-1}(q+p).
} 
for the gluon-matter-matter vertex $\gamma_{\mu\alpha\beta}$ holds.
Therefore this graph gives a contribution $c_i^{(b)}$ to the relevant
coefficients plus a contribution of order 
$O(\Lambda^2/\Lambda_0^2,p^2/\Lambda_0^2)$ to the irrelevant coefficients.
The tadpole graph (figure 3c) is 
momentum independent and gives solely a contribution to the coefficients
$c^{(c)}_3$ and $c_4^{(c)}$.
The important point is that it is possible to recover the transversality of
the self-energy at the
scale $\Lambda$ provided that we choose the ultraviolet
boundary conditions on the
relevant couplings associated to non-transverse terms as
\formula{fine-tuning}
{p^\mu\Pi_{\mu\nu,rel}^{(1)}(p,n;\Lambda_0)\equiv p^\mu F_{\mu\nu,rel}^{reg}
(p,n;\Lambda_0)-p^\mu F_{\mu\nu,rel}^{reg}(p,n;\Lambda).
}
This is the fine-tuning procedure. In this way the non-transverse terms are
vanishing when the ultraviolet cutoff $\Lambda_0$ is removed and therefore
gauge-invariance is recovered at the end.

Using the classical Ward identities \rif{W.cl.1}\--\rif{W.cl.3} one could apply
the same arguments to all the remaining one-loop Ward-identities. However 
we prefer to give a general proof based on the Quantum Action Principle.

\section{General proof of gauge-invariance}

The effects of a non-gauge-invariant regularization
can be studied in the formalism of the modified Ward identities. In appendix 
B we prove that the variation of the 
effective action under linear transformation when a generic cutoff 
function $R_\Lambda(q)$ is employed has the form
\formula{mWI}
{\delta\Pi(\Phi;\Lambda)=
i\hbar\STr R_\Lambda L(R_\Lambda+\Pi_{\tilde\Phi\Phi})^{-1}-\Delta_\Gamma
}
where $\Delta_\Gamma$ is local (this is the content of the
Quantum Action Principle). 
Therefore the condition $\Delta_\Gamma=0$ 
is equivalent to a system of equations for the spurious (i.e. associated to
non-invariant operators)
relevant couplings which can be solved 
in perturbation theory. These are the general fine-tuning conditions.
In particular we can introduce
a class of regular cutoff terms which limit is our specific cutoff, 
\formula{Q.UV}
{R^{\Lambda_0}_{\Lambda,\mu\nu}(q^2)=
\bar K_{0\Lambda_0}(q^2)\left(\Lambda^2
g_{\mu\nu}+\frac1{\xi_2}\frac{n_\mu n_\nu}{n^2}\right).
}
A class as \rif{Q.UV} has been implicitly used in 
the one-loop analysis of the gluon self-energy 
transversality given in the section 5 since it corresponds to 
a propagator exponentially
dumped in the ultraviolet such as \rif{UV.reg}. 

If for notational simplicity we work in the pure Yang-Mills case
(the extension to a vector-like coupled matter is trivial)
we can rewrite \rif{mWI} in  a more explicit notation as
\formula{mWI.explicit}
{\W_f\Pi=\int_{x,y}iR^{\Lambda_0}_{\Lambda,\mu\nu}(-\partial_x^2)
\delta(x-y)f(x)\cdot<0|T A^\mu(x)\times A^\nu(y)|0>_{c,J},
} 
where $<\dots >_{c,J}$ 
denotes the connected correlation function in presence of
the non-zero source $J_\mu=-\dd\Gamma{A^\mu}$. It is clear from this formula
that when the ultraviolet cutoff is removed ({\it after}
the correct subtractions in the proper vertices has been done) 
the right hand side is identically zero simply due to the
antisymmetry of the structure constants.
Therefore even in this case we can prove
\formula{g.i.}
{\delta\Pi(\Phi;\Lambda)=0.
}
As a matter of fact, all the results reported in this paper can be generalized,
at least at the perturbative level, in a way independent of the ultraviolet
regularization, provided that the fine tuning problem is solved. 
The new point of this paper is the fact that actually it is possible 
to avoid the fine-tuning procedure by adopting a gauge-consistent 
regularization, the simplest being dimensional regularization.

\section{The simple proof of gauge-invariance}

If we adopt a gauge-consistent intermediate regularization then 
gauge-invariance can be directly proved from the functional form of the ERGE
\footnote{One could give a third proof of gauge-invariance by
deducing \rif{g.i.} by standard manipulations in the functional integral.
Actually, this is the less rigorous but simpler approach.}
\cite{paper.I,Wetterich1}
\formula{sERGE}
{\dot\Pi(\Phi;\Lambda)=I(\Phi;\Lambda)=\left[-
\frac i2\Str\dot Q_\Lambda\left(
Q_\Lambda+\Pi_{\tilde\Phi\Phi}\right)^{-1}\right]_{reg-inv},
} 
where we have defined
\formula{Pi..}
{\Pi_{\tilde\Phi\Phi}=\dd{}\Phi\ddes{}{\Phi}\Pi(\Phi;\Lambda).
}
The logic of the proof is the following \cite{paper.I}.
\begin{enumerate} 
\item We suppose that
the functional $\Pi(\Phi;\Lambda)$  is gauge-invariant 
at some initial scale $\bar\Lambda$,
\formulaX
{\W_f\Pi(\Phi;\bar\Lambda)=0.
}
\item We observe that in this
hypothesis even the functional $I(\Phi;\bar\Lambda)$ is gauge-invariant,

\formulaX
{\W_f I(\Phi;\bar\Lambda)=0.
}

\item Therefore
the evolution equation is gauge-invariant and, as a consequence, 
the Ward identities are satisfied to any $\Lambda$.
\end{enumerate}

Here we give a very elegant proof, valid for generic linear 
symmetries\footnote{Obviously this proof also applies to the Abelian case in 
covariant gauges
and may be seen as an alternative analysis with respect to the
diagrammatic proof of \cite{paper.I}.}, i.e.
for infinitesimal $\Lambda-$independent field transformations of the kind
\formula{tr.lin}
{\delta\Phi^A=L^A_B\Phi^B+\ell^A
}
such as the classical action is invariant
\formula{sim.lin}
{\delta S_{CL}(\Phi;\Lambda)=\left(L\Phi+\ell\right)^A\dd {S_{CL}}{\Phi^A}=0.
}
For instance in a pure Yang-Mills theory (but the same property holds even when
the coupling with the matter is considered) the gauge symmetry
$\delta A_\mu=\W_f A_\mu=-f\times A_\mu+\partial_\mu f$ is linear with
\formulaX
{L^A_B=-f^a_b(x)\delta(x-y)\delta^\mu_\nu,\quad 
f^a_b(x)\equiv f^c(x) (\tau_c)^a_b,\quad 
\ell^A=\partial^\mu f^a(x).
}
In general the quantization procedure can break the symmetry. However, 
if the cutoff function $Q_\Lambda$
has the exceptional property (which holds in our case)
\formula{comm}
{Q_\Lambda \tilde L+L Q_\Lambda=0,}
then the breaking term is linear in quantum fields
\formula{W.tree}
{\delta \Gamma^{(0)}(\Phi;\Lambda)=\tilde\ell Q_\Lambda\Phi
}
and the evolution equation is gauge invariant.
To prove this statement we use
the transformation law of $\Pi_{\tilde\Phi\Phi}(\Phi;\bar\Lambda)$ under 
finite transformations
\formula{finite}
{\Phi'=e^L\Phi+\ell} 
(in geometric language, these 
transformations 
are affine isometries with respect to the metric in the fields space 
defined by $Q_{\Lambda,AB}$) which is simply
\formulaX
{\Pi_{\tilde\Phi\Phi}(\Phi';\bar\Lambda)=e^{L}\Pi_{\tilde\Phi\Phi}
(\Phi;\bar\Lambda) e^{\tilde L}.}
By using the explicit form of $I(\Phi;\bar\Lambda)$ and 
the invariance property
of $Q_\Lambda$, which follows from \rif{comm}
\formulaX
{e^{L}Q_\Lambda e^{\tilde L}=Q_\Lambda,}
one immediately prove the gauge invariance of the evolution equation at the
scale $\bar\Lambda$
\formulaX
{I(\Phi';\bar\Lambda)=I(\Phi;\bar\Lambda),
}
therefore at all scales.

\section{Gauge-dependence and unitarity}

In order to give a rigorous status to the axial gauge formulation, 
one should prove the following statements:
\begin{enumerate}
\item Physical amplitudes are consistent with unitarity.
\item Physical quantities are independent of the gauge vector $n_\mu$.
\end{enumerate}
The fulfillment of these properties is non-trivial 
in the usual axial gauge formulation, due to the difficult problem 
of spurious divergences. In particular a canonical 
analysis shows that the unitarity issue is very subtle \cite{Naka}.
In a generic Wilsonian framework \cite{Litim}
the spurious divergences are avoided, but 
there is an explicit unitarity breaking at $\Lambda\neq0$ due to the
breaking of Ward identities.
Fortunately, the situation is better in our formalism and we can
easily prove the consistency with unitarity for any finite $\Lambda
\neq0$ \cite{Soldati.unit}. However, we will lose covariance for any
$\Lambda\neq0$.

To show unitarity, we have to prove that our propagator \rif{D.axial}
is consistent with the Landau-Cutkosky rules.
Let us consider an orthonormal basis $e^{(\mu)}_\lambda(p,n)$ satisfying
\formula{ortho}
{e^{(\mu)}_\lambda e^{(\nu)*}_\rho g^{\lambda\rho}=g^{\mu\nu},
\quad e^{(\mu)}_\lambda e^{(\nu)*}_\rho g_{\mu\nu}=g_{\lambda\rho}
}
and let us take the vector $P_\mu=p_\mu-n_\mu pn/n^2$ which is on-shell 
time-like, i.e. $P^2>0$ for $p^2=\Lambda^2$. If we define
\formulaX
{e^{(0)}_\mu=P_\mu/\sqrt{P^2},\quad e^{(3)}_\mu=n_\mu/\sqrt{-n^2},
} 
then the property
\formula{sum.pol}
{\sum_{i=1}^2e^{(i)}_\mu e^{(i)*}_\nu=-g_{\mu\nu}+\frac{-pn(n_\mu p_\nu+p_\mu
n_\nu)+n^2p_\mu p_\nu+p^2 n_\nu n_\nu
}{p^2n^2-(p n)^2}
}
holds. As a consequence we can write
\formula{Im.D}
{\mbox{Im} D_{\Lambda,\mu\nu}(p)=\pi 
\delta(p^2-\Lambda^2)\theta(p_0)\sum_{i=1}^2e^{(i)}_\mu e^{(i)*}_\nu
}
and this assures the consistency with the optical theorem and unitarity.
Therefore apparently
we have constructed an unitary massive renormalizable non-Abelian 
gauge theory. Nevertheless, this theory is unphysical at $\Lambda\neq0$ 
due to an incurable breaking of Lorentz covariance.

To understand this point we consider, as a specific example, the 
gauge-invariant quantity 
\formulaX
{G(p,n)=\int_x e^{ipx}<0|T [F_{\mu\nu}F^{\mu\nu}](x)
[F_{\mu\nu}F^{\mu\nu}](0)|0>.
}
At lowest order in perturbation theory this quantity is given by a
Feynman diagram containing the transverse part of the axial gauge 
propagator,
\formulona{D.T.axial}
{-D^T_{\mu\nu}(p;\Lambda)&=-t_{\mu\rho}(p)D^{\rho\sigma}(p;\Lambda)
t_{\sigma\nu}(p)\\
&=\frac{t_{\mu\nu}(p)}{p^2-\Lambda^2+i\ep}+\frac{\Lambda^2
N_\mu N_\nu}{((p\cdot n)^2-n^2\Lambda^2)(p^2-\Lambda^2+i\ep)},
}
with $N_\mu= n_\mu-p_\mu pn/p^2.$
From the explicit 
formula one sees that at $\Lambda=0$ the $n_\mu-$dependence cancels:
however for $\Lambda\neq0$ even gauge-invariant observables {\it does} 
depend on the
quantization direction $n_\mu$. Nevertheless, in the
ultraviolet region the $n_\mu$-dependence, i.e.
the explicit Lorentz-covariance breaking term, is suppressed. 
Therefore the Wilsonian approach can give reasonable results 
even when $\Lambda$ is not exactly zero. In other words we expect
the $\Lambda-$dependence be analytical for 
a certain physical observables in some momentum range and for reasonable
approximation schemes.
This point is of crucial importance because 
in numerical simulations $\Lambda$ is never completely removed.
However we notice that there are situations where the na\"\i ve 
$\Lambda\to0$ limit does not exists; and the subtle problems connected
with this limit will be discussed in a forthcoming paper \cite{paper.III}.

Here we are more interested in the Lorentz-covariance issue.
We want to provide a machinery to control the Lorentz-covariance 
breaking terms. The method we use is based on BRST techniques 
and directly inspired to analysis of \cite{Gaigg}. 

Consider first the pure Yang-Mills case (the analysis trivially generalize
to a vector-like matter coupling) at $\Lambda=0$.
As it is well known \cite{Bassetto}, even in the axial gauge framework
one can introduce the ghosts $C^a$ and $\bar C^a$ and the auxiliary field
$\lambda^a$ with BRST transformations
\formula{BRST.NC}
{sA=D C,\quad sC=-\frac12g C\times C,\quad s\bar C=\lambda,\quad s\lambda=0
}
and one can take as tree level action\footnote{In
general axial gauges one adds the BRST-trivial operator 
$s\left(\bar C\cdot n  A-\frac{\xi_2}2\bar C\cdot\lambda\right)=
n_\mu A^\mu\cdot\lambda-\bar C n^\mu D_\mu C-\frac{\xi_2}2 \lambda^2$.}
\formula{n-coupling}
{S_{BRST}=S_{CL}(A)+\int_x s\left(\bar C n_\mu A^\mu\right)=
S_{CL}(A)+\int_x n^\mu\left
(A_\mu\lambda-\bar C D_\mu C\right).
}
Now it 
is clear from \rif{n-coupling} that the quantization direction $n_\mu$
cannot influence physical quantities, since it couples with a BRST-trivial
operator. 
However, if the infrared cutoff term $\frac12\Lambda^2A_\mu\cdot A^\mu$ is
considered, the BRST symmetry is explicitly broken,
$$
s\left(S_{BRST}+\int_x\frac12\Lambda^2A_\mu\cdot A^\mu\right)=
\int_x\Lambda^2A_\mu\partial^\mu C\neq0.
$$
Nevertheless, there is a way to control the BRST breaking term and the
$n_\mu$-dependence by extending
the BRST-symmetry in a suitable way. To this aim we introduce two 
Grassmann constants $\bar\rho$ and $\varphi_\mu$ with mass dimension and
Grassmann number respectively 2,-1 and 0,1 (see table 1) and 
extended BRST-transformation
\formula{s.ext}
{s_E\ \bar\rho=\Lambda^2,\quad s_E\ \varphi_\mu=0;}
moreover we give to the gauge vector the transformation
\formula{s.n}
{s_E\ n_\mu=\varphi_\mu;}
on the quantum fields $s_E$ exactly coincides with $s$. 
Now the extended action
\formulonaX
{S_{EXT}(\Phi;\Lambda^2,\varphi_\mu,\bar\rho)&=
S_{CL}(A)+s_E\left(\bar C\cdot 
n_\mu A^\mu+\frac12\bar\rho A_\mu\cdot A^\mu\right)\\
&=S_{CL}(A)+\int_x nA\cdot\lambda-\bar C\cdot n^\mu D_\mu C-\bar C\cdot
\varphi_\mu A^\mu\\
&\quad+\frac12\Lambda^2 A\cdot A-\bar\rho A^\mu\cdot \partial_\mu C}
is invariant under the extended BRST-symmetry and all the usual techniques
can be used to lift this symmetry to the quantum level in the form of
a generalized Slavnov-Taylor equation which reads
\tabellaX{Pure Yang-Mills theory.}{||c|c|c||}
{\hline Field or parameter & Mass dimension & Ghost number\\
\hline\hline
$A$ &  1 &  0\\
\hline
$C$ & 0 & 1\\
\hline
$\bar C$ & 3 & -1\\
\hline
$\lambda$ & 3 & 0\\
\hline
$n_\mu$ &0& 0\\
\hline
$\varphi_\mu$ &0& 1\\
\hline
$\bar\rho$ &2 & -1\\
\hline 
$A^*$ & 3 & -1\\
\hline
$C^*$ & 4 & -2\\
\hline} 

\formula{ST.ax}
{\int_x\left(\dd{\Gamma^{(0)}}{A^*}\cdot\dd{\Gamma^{(0)}}{A}+
\dd{\Gamma^{(0)}}{C^*}\cdot\dd{\Gamma^{(0)}}{C}+\lambda\cdot\dd{\Gamma^{(0)}}
{\bar C}\right)+\varphi_\mu
\dede{\Gamma^{(0)}}{n_\mu}+\Lambda^2\dede{\Gamma^{(0)}}{\bar\rho}=0.
} 
In \rif{ST.ax}, as usual, we have introduced two external sources $A_\mu^*$ 
and $C^*$ coupled to
the non-linear variation $sA_\mu$ and $sC$, respectively.
The perturbative machinery allows to extend this identity to all
orders in the loop-wise expansion\footnote{
In the Wilsonian point of view, one defines 
$$\int_x\left(\dd{\Gamma}{A^*}\cdot\dd{\Gamma}{A}+
\dd{\Gamma}{C^*}\cdot\dd{\Gamma}C+\lambda\cdot\dd{\Gamma}
{\bar C}\right)+\varphi_\mu
\dede{\Gamma}{n_\mu}+\Lambda^2\dede{\Gamma}{\bar\rho}=\Delta_\Gamma
$$
and checks the consistency with the evolution equation, by showing that 
$\Delta_\Gamma=0 \Longrightarrow \dot\Delta_\Gamma=0$. 
Technically is simpler to study the 
functional $\Delta_W(J)=\Delta_\Gamma(\Phi)$, as done in appendix B. 
In the proof the identity
$-\dede W{\bar\rho}=\int_x\dd{W}{A_\mu^* \delta J^\mu}+i
\dd W{A_\mu^*}\dd W{J^\mu}$ is helpful.} since no anomalies are generated.
Now by deriving with respect to $\varphi_\mu$ and by 
putting $\varphi_\mu=0$
one obtains the functional dependence on the gauge vector $n_\mu$.
In particular in the $\Lambda\to0$ limit one proves the gauge-independence
of physical quantities \cite{Gaigg}.
Moreover in $\Lambda\to\infty$ limit, when the effective action becomes
local and has the {\it same} form of the (extended) BRST-action, 
the symmetry forbids the presence
of the $h(\Lambda)$ term introduced in \rif{general.form}. In other words, 
$h^{(\ell)}(\Lambda)$ must be subleading with respect to the other
relevant couplings and in particular at one-loop is finite (actually 
$h^{(1)}(\Lambda)\equiv0$ since the boundary condition
$h(\Lambda_R)=0$ is imposed).

The important point we wish to stress here, is that 
the axial BRST symmetry is much simpler
than the covariant usual one: this is due to the fact that the additional
fields $(\lambda,\bar C, C)$ we have introduced {\it decouple} completely 
from the theory. 
This is due respectively to the constraint equation
\formula{lambda.eq}
{\dd{\Gamma^{(0)}}\lambda=\Delta_\lambda^{(0)}=n^\mu A_\mu,
}
the ghost equation
\formula{ghost.eq}
{\dd{\Gamma^{(0)}}{\bar C}-gC\times\dd{\Gamma^{(0)}}\lambda
=\Delta_{\bar C}^{(0)}=-n^\mu\partial_\mu C-\varphi^\mu A_\mu}
and finally the anti-ghost equation
\formula{anti.ghost.eq}
{\dd{\Gamma^{(0)}} C-g\bar C\times\dd{\Gamma^{(0)}}\lambda=\Delta_C^{(0)}}
with 
\formulaX
{\Delta_C^{(0)}=-n^\mu\partial_\mu\bar C-
\bar\rho\partial_\mu A^\mu+\partial^\mu A_\mu^*- gA^*\times A-
g C^*\times C.
}
These functional identities correspond to (non-anomalous) linearly 
broken linear symmetries and can be extended to all orders in perturbation 
theory. In particular the right hand side does not renormalize.
Notice that this fact is independent of the specific adopted renormalization 
procedure, since the general analysis described in appendix B and based on
the QAP applies.
Therefore the crucial features of the axial gauge
formulation are maintained even at $\Lambda\neq0$. 

\section{Beta function computation}

As we explained in full detail in \cite{paper.I} one can extract
from the relevant part of the Wilson Renormalization Group Equation
the Callan-Symanzik renormalization group functions i.e. the beta
function and the anomalous dimension. In this section we present,
as a non-trivial consistency check of the formalism, a one-loop 
computation of the QCD beta function. 
Here we will restrict to the pure gauge part,
since the contribution from the matter fields is exactly the same as
in covariant gauge. This example will be worked out in detail, since the 
computation is quite different with respect to
more standard analysis.

As it is well known, one of the technical
advantages of the axial gauge formulation is the fact that
the beta function can be directly computed  from the gluon self-energy,
thanks to the QED-like identity \rif{QED.like}
which relates the beta function to 
the anomalous dimension $\gamma_A(g)$,
\formula{beta.gamma}
{\beta(g)=g\gamma_A(g),\quad \gamma_A(g)=-\frac12\frac{\dot Z_A}{Z_A}.
}
Taking in account the fact that, as a consequence
of the generalizated BRST symmetry, the relation $\dot h^{(1)}(\Lambda)=0$
holds (this can also be checked with an explicit computation), we 
can directly 
extract the wave function $\dot Z_A^{(1)}$ from the gluon self-energy.
By using the choice $n^\mu=(0,0,0,1)$ the relevant integral 
comes from the 33-component of the self-energy,
\formula{ZA}
{\dot Z_A^{(1)}=\partial_{\bar p^2}\dot\Pi^{(1)}_{33}|_{p=0},\quad
\bar p=(p^0,p^1,p^2).
}
Due to the simple form of the (Euclidean) gluon propagator 
\formulona{prop.eucl}
{D_{\Lambda,33}&=D_{\Lambda,3i}=D_{\Lambda,i3}=0,\quad i=0,1,2\\
D_{\Lambda,ij}&=\frac{\delta_{ij}}{p^2+\Lambda^2}
+\frac{p_i p_j}
{(p^2+\Lambda^2)(p_3^2+\Lambda^2)}
}
and the explicit form of the trilinear vertex,
there is a big simplification in the integral defining $\dot Z_A^{(1)}$
with respect to the general integral \rif{1.loop.gluon}. Moreover,
by using the trick reported in appendix C equation 
\rif{trick} one can reduce this
integral to an even simpler integral in $\bar q$ evaluated at $\Lambda=0$
which explicit form is
\formula{beta.int}
{\dot Z_A^{(1)}=-g^2N_c\int_{\bar q}
\frac43\ \frac{3\ q_3^6+2\ q_3^2\ \bar q^4+\bar q^6}{(\bar q^2+
q_3^2)^4\ q_3\ \pi}.
}
This integral can be computed by using equation
\rif{dk3} and finally the expected result
\formula{beta.1}
{\beta^{(1)}(g)=-g\dot Z_A^{(1)}/2=-\frac1{16\pi^2}\frac{11}3N_C\ g^3
}
is obtained.

We remark that the computation is actually 
{\it independent} of the cutoff function choice. Therefore 
the same result can be obtained even in other approaches, when generic
cutoffs are employed. This feature is due to two reasons:
\begin{enumerate}
\item The identity 
$\beta^{(1)}=g\gamma_A^{(1)}$ holds even if the gauge symmetry is broken
(but {\it only} at one-loop). This a consequence of the Quantum Action
Principle.
\item  The relation \rif{trick} in appendix C
allows to relate an integral apparently
dependent on the infrared cutoff $\Lambda$ to an universal integral 
independent of the cutoff function choice.
\end{enumerate}

\section{Conclusions}

In this paper we have solved the problem of giving an explicitly 
gauge and unitarity consistent Wilsonian formulation of non-Abelian gauge
theories. The price to pay is the lost of covariance, i.e. there is 
an unphysical dependence on the gauge vector $n_\mu$ for $\Lambda\neq0$. 
However this dependence vanishes in the $\Lambda\to0$ limit for 
physical quantities (assuming we consider a quantity such as the 
$\Lambda\to0$ limit exists) whereas for finite 
$\Lambda$ is controlled by a simple generalized Slavnov-Taylor identity.
Therefore there is a strong progress with respect to generic Wilsonian 
procedures,
where gauge-invariance and unitarity are lost and the 
gauge-dependence is much more difficult to study. 

In the approach we presented, one not only maintains all 
the expected advantages both of the non-covariant gauge
formulation (simple Ward identities and decoupling of the ghosts) 
and of the Wilsonian viewpoint
(simple understanding of the renormalizability issue and of
the relation with the field theory renormalization group),
but there are also additional bonus: in particular
spurious divergences are automatically avoided and 
we have efficient methods to perturbatively compute 
beta functions and anomalous dimensions.
However, the greatest  potentiality of this approach relies on the possibily
of starting a non-perturbative analysis via suitable
{\it gauge-invariant} truncations, 
thus solving a major problem of the usual
ERGE approach to gauge theories \cite{Ellw.rel,Jungnickel}.

An important 
point which has not been addressed in this paper, is
the question of the infrared limit $\Lambda\to0$. This point is very
delicate. As a matter of fact, in the usual formulation with the CPV 
prescription, various inconsistencies have been found 
\cite{Soldati}. 
These problems of principle of the axial gauge choice will stay 
also in the Wilsonian approach in a disguised form, as
we will discuss in \cite{paper.III}. Probably the simpler way to avoid these 
problems is to switch to planar or light-cone gauges, which are safe 
in the usual formalism. This will be the subject of 
\cite{paper.III}.
However, the main results of this paper i.e. the consistency
with Ward identities and the control of the Lorentz-covariance
breaking (i.e. the $n_\mu-$dependence) straightforwardly generalize
to all non-covariant gauges.

Moreover, it is straightforward to extend the part of this work concerning
the (extended) BRST symmetry to the covariant gauge case. Actually it is
possible to control the BRST breaking mass term and the
gauge-dependence issue by using the same tricks we have used 
in this paper. However in this case unitarity is lost at $\Lambda\neq0$,
ghosts do not decouple,
and the analysis of Slavnov-Taylor identities
is less simple. The obvious
advantage is that the manifest covariance is maintained.
In general the problem of how non-linear symmetries are treated 
in our version of the Wilson Renormalization Group is an interesting issue
which should be investigated in the future. The only non-trivial point is 
that the
definition of relevant couplings in terms of zero-momentum subtractions is
inconsistent with non-linear symmetries and cannot be maintained. Instead
some minimal subtraction similar to the dimensional regularization procedure
must be invoked.

\vskip .5cm
{\large\bf Acknowledgements}\\

I am indebted to A. Bassetto, U. Ellwanger, T. Morris and R. Soldati 
for e-mail correspondence. 
I acknowledge LPTHE for kind hospitality and in particular 
H. J. De Vega and J. Salgado.
This work has been supported from foundation ``Aldo Gini'' and INFN, 
Gruppo Collegato di Parma.

\appendix

\section{Sketch of the renormalizability proof}

For sake of completeness, in this appendix we briefly comment about the
renormalizability property in the axial gauge framework.
The point is that 
the Wilsonian renormalizability proof stated in \cite{paper.I} for a
large class of cutoff functions can be
extended to the present case. 
Here we simply sketch the 
argument, a rigorous proof can be easily given by following the
lines described here and the work done in \cite{paper.I}.

Technically speaking, 
the great advantage of the Wilsonian formulation is the fact that,
once one has proved that the integrated evolution equation is well defined when
the ultraviolet cutoff is removed at the first order of iteration, 
then this property is immediately transported to all orders 
in perturbation theory. In other words, no analysis 
of overlapping divergences is required \cite{Polchinski} to prove
that the proper vertices are well defined in the $\Lambda_0\to\infty$
limit.

In the case at hand, one sees
that all the one-loop Feynman diagrams in the integrated evolution equation
are convergent by direct inspection. This is straightforward since
in the renormalizability proof
the infrared cutoff $\Lambda$ is different from zero and therefore
the spurious divergences are automatically avoided.
In particular for $n^\mu=(0,0,0,1)$, 
one can split the four dimensional integration
$\int_q$ as $\int_{q^3}\int_{\bar q}$. All the three-dimensional integrals 
are made finite by using the natural prescription reported in appendix C;
therefore one must only control the $q_3$ integrals which are
convergent for dimensional analysis, due to the momentum subtractions.

In a more conventional language the renormalizability is expected since 
in the axial gauge formalism there is a power counting criterium
and the ultraviolet divergences are local polynomials in the momenta 
\cite{Bassetto}.

\section{Formal proof of the mWI and relation with the QAP}

In various part of the text we have used some results from the
Quantum Action Principle (QAP). For completeness, here we review its
proof in the Wilson Renormalization Group formulation, by focusing
in particular on the analysis of (possibly linearly broken) 
linear symmetries. 
The source of this kind of rigorous proofs can be found in the work of
Becchi \cite{Becchi}.  A translation in the cutoff action formalism is given 
in \cite{Datta} (see also \cite{Ellw}). Both references 
consider the non-linear symmetries, but the analysis 
is essentially the same also  for linear symmetries.\\ 
Let us consider the general modified Ward identity (mWI)
\formula{Delta.gen}
{-\delta\Pi+i\hbar STr R_\Lambda L(R_\Lambda+\Pi_{\tilde\Phi\Phi})^{-1}=
\Delta_\Gamma,}
where $\hbar$ has been explicited
since our analysis will be perturbative in the loop-wise expansion.
The QAP \cite{QAP} says that if $\Delta_\Gamma^{(\ell')}=0$ for $\ell'=0,\dots,
\ell$, then $\Delta_\Gamma^{(\ell+1)}$ is local,
i.e. contains a finite number of relevant terms. 
Under certain conditions, i.e. in absence of anomalies, 
this fact assures that it is possible to impose
\formula{Delta0}
{\Delta^{(\ell)}_\Gamma(\Phi;\Lambda)=0} 
to all orders.
Technically, the proof proceeds by showing that the functional identity
\rif{Delta0} is consistent with the evolution equation. 
Therefore if it holds at the initial scale $\bar\Lambda$ it holds 
at all scales. The proof 
is independent of the used cutoff function $R_\Lambda(q)$, 
which can be at large extent generic.

To prove the statement it is convenient
to study the functional
\formula{DeltaW.}
{\Delta_W(J;\Lambda)=\left(\tilde J+\ddes WJ R_\Lambda\right)
\left(L\dd W{J}+\ell\right)-
i\hbar\Str R_\Lambda L\dd {}J\ddes WJ.
}

Notice that $\Delta_\Gamma(\Phi;\Lambda)=\Delta_W(J;\Lambda)$ 
is zero {\it iff} $\Delta_W(J;\Lambda)$ is zero.
To prove this latter fact we use the evolution equation of the 
$W(J;\Lambda)$ functional \cite{paper.I}
\formula{W.ERGE}
{\dot W=-\frac i2\STr\dot R_\Lambda\left(
\hbar\dd{}J \ddes WJ+i\dd W J \ddes WJ\right).
}
By a lengthy but straightforward computation one
obtains the evolution equations of $\Delta_W$ which is {\it linear} of the kind
\formula{linear.eq}
{\dot\Delta_W=M_W\Delta_W}
where $M_W$ is the functional operator
\formula{def.M}
{M_W=-\frac i2 \STr\dot R_\Lambda\left(\hbar
\dd{}J \ddes{}J+i\dd W J\ddes{}{J}+i\dd{} J\ddes W{J}\right).}
Therefore if $\Delta_W$ is zero at some scale $\bar\Lambda$, it is identically
zero and the functional identity \rif{Delta0} holds.
Equation \rif{linear.eq} has been introduced and solved 
in \cite{Becchi}; however it is more convenient to consider the equivalent
identity \cite{Datta}
\formula{linear.fund}
{\LdL\Delta_\Gamma=\hbar M_\Gamma\Delta_\Gamma} 
where
\formula{M.Gamma}
{M_\Gamma=\frac i2\STr
(R_\Lambda+\Pi_{\tilde\Phi\Phi})^{-1}\dot R_\Lambda
(R_\Lambda+\Pi_{\tilde\Phi\Phi})^{-1}
\dd{}{\Phi}\ddes{}\Phi.} 
Equation \rif{linear.eq} is simpler since can be solved recursively in 
$\hbar$  and can be used to give an easy proof of 
the QAP in the Wilsonian formalism.
We remind that the QAP is 
the basic building block of the algebraic renormalization 
program \cite{Piguet} and is conceptually very important: in particular
a renormalization scheme  should be considered viable only if it is consistent
with this principle\footnote{For instance, we remind that
dimensional reduction in supersymmetric
theories is a largely used but inconsistent renormalization scheme 
\cite{Bonneau}.}. Fortunately, the proof of the QAP is very simple in the
Wilson Renormalization Group approach, at least at the perturbative level.
In fact from \rif{linear.fund}, by using the fact that
$\Delta_\Gamma^{(0)}$ vanishes (i.e. the
symmetry holds at the classical level) one obtains that 
$\Delta^{(1)}_\Gamma$ is $\Lambda-$independent. Therefore it
is equal to its value at the ultraviolet
scale $\Lambda_{UV}>>\Lambda_R$. 
But at this scale $\Delta_\Gamma^{(1)}$ is local for dimensional
reasons (the irrelevant parts are suppressed as inverse powers 
of $\Lambda_{UV}$) therefore the QAP holds at one loop: 
the breaking term is local.
Now, may be possible to impose $\Delta_\Gamma^{(1)}=0$ 
at the scale $\Lambda_{UV}$,
depending on the algebraic structure of the theory, i.e. on the
study of the Wess-Zumino consistency conditions \cite{Piguet}.
If the structure is trivial (i.e. there are no anomalies)
we can impose $\Delta^{(1)}_\Gamma=0$. This involves a non-trivial
fine-tuning of spurious couplings in terms of the physical couplings
which explicit form is given by the one-loop relevant part of equation
\rif{Delta.gen}.
At this point one can iterate the argument and use $\Delta_\Gamma^{(1)}=0$ and
\rif{linear.fund} to
prove that $\Delta_\Gamma^{(2)}$ is $\Lambda-$independent and therefore
local: thus by a new fine-tuning one can impose $\Delta_\Gamma^{(2)}=0$.
The argument obviously generalize to all orders in the loop-wise expansion. 
As a consequence the mWI \rif{Delta.gen} holds for any $\Lambda$. \\
Notice that exactly the {\it same} analysis holds in the
case of theories which are linearly broken at three level. In fact, due
to the explicit form of the $M$ operator, which contains second order
functional derivatives in the fields, one directly see that all
the perturbative corrections $\Delta_\Gamma^{(\ell)},\ \ell\geq1,$ 
identically vanishes. Therefore we recover in this context
the well known results that linear breaking terms does not renormalize.
In particular this remark applies to the functional identities \rif{lambda.eq},
\rif{ghost.eq} and \rif{anti.ghost.eq} in the text, as well as to the Ward 
identities \rif{mWI}, and is completely general, i.e. independent on the
used renormalization procedure, provided that the QAP holds and the
theory is anomaly free.

In general the previous analysis is required whenever one is 
interested in theories where the ERGE is inconsistent with the classical
symmetries. However in various cases it is impossible to impose 
$\Delta_\Gamma=0$. For instance in chiral gauge theories\footnote{
In our formalism the ERGE is inconsistent with chiral symmetry 
due to the spinor ``mass'' term $i\Lambda\bar\psi\gamma_5\psi$.}
the right hand side of the mWI \rif{Delta.gen} is non-zero and one could 
compute the chiral anomaly by following the lines of \cite{B.Vian}. 
Obviously one obtains the same result since the coefficient
of the anomaly is independent of the cutoff function choice.

The important 
point of this paper is the fact that we were able to avoid 
such a non-trivial analysis in non-anomalous theories like QCD. In fact, 
by using the Wilsonian renormalization procedure we have presented, 
supplemented with a gauge-consistent intermediate regularization,
the tree-level Ward identities are never broken and therefore $\W_f\Pi=0$ 
identically holds, {\it without fine-tuning}.

\section{Technical remarks}

In this appendix we collect some formulae helpful in perturbative computations.

First we present an useful trick allowing to compute with a little effort
dimensionless coefficients of the kind
\formula{RG.coeff}
{c=\int_{q_E}\LdL F(q_E;\Lambda)}
where $F(q_E;\Lambda)$ ha mass dimension $-4$ and is obtained from
some Feynman diagram at zero momentum, possibly 
derived with respect to the external momenta.
Typically $c$ is a renormalization group coefficient, in particular the
one-loop beta function coefficient. 
The trick we present\footnote{A similar 
relation holds in covariant computations and has been proved in
the appendix of \cite{BS}.} allows to rewrite equation \rif{RG.coeff}
in the form
\formula{trick}
{c=-\int_{\bar q} \frac{q_3}\pi F(\bar q,q_3;0)
}
which is simpler since the integration in $q_3$ and the
$\Lambda-$dependence disappeared. Notice that
since $c$ is dimensionless, the $q_3$-dependence is apparent and
cancels, i.e. $\int_{\bar q}F(\bar q,q_3;0)\sim1/q_3$.
The derivation of equation \rif{trick} is based on the relation 
$\LdL f(q_3/\Lambda)=-q_3\partial_{q_3} f(q_3/\Lambda)$ from which
one obtains
$$
\int_{q_E}\LdL F=2\int_0^\infty\frac{dq_3}{2\pi q_3}\LdL \int_{\bar q}
q_3F(\bar q,q_3;\Lambda)=-\int_{\bar q}\frac{q_3}\pi F(\bar q,q_3;0),
$$
where the property $F(\bar q,q_3;\Lambda\to\infty)=0$ has been used.

The tree-dimensional integrals appearing in \rif{trick} can be evalued
with standard formulae. 

\begin{itemize}

\item Formulae for three-dimensional angular averages.
 
If $m=2n$ is even, we have
\formula{int.ang}
{<(\vec q\cdot \vec p)^m>_{\Omega_3}
\equiv\frac1{4\pi}\int_0^\pi d\theta\int_0^{2\pi} 
d\phi \sin\theta (q p \cos\theta)^{2n}=\frac{(q p)^{2n}}{2n+1},
} 
otherwise, if $m=2n+1$ is odd, we have
\formula{int.ang2}
{<(\vec q\cdot\vec p )^m>_{\Omega_3}\equiv0.
}
\item Formulae for the momentum integration.
 
Let be $N>0,\ M\geq0$  integers and $A>0$ 
a $k-$independent real number; for $N>M+3/2$ the relation
\formula{dk3}
{\int\frac{d^3k}{(2\pi)^3}\frac{k^{2M}}{(k^2+A)^N}=\frac1{4\pi^2}
\frac{B(N-M-3/2,M+3/2)}{A^{N-M-3/2}},
} 
where $B(a,b)=\Gamma(a)\Gamma(b)/\Gamma(a+b)$
is the Euler's Beta function, holds. The relation \rif{dk3}
can be continued to $N<M+3/2$, by implicitly defining
a (natural) prescription for three-dimensional 
divergent integrals\footnote{Such a prescription
does not work in even dimensions of space time, since in this case
the Beta function develops poles.}. 
However in this case the integral is positive only if 
$-2n<N-M-3/2<-2n+1$, with
$n=1,2,3,\dots$. This observation explains the origin of the negative
sign for the beta function in our computation.

\end{itemize}


\end{document}